\documentclass[10pt]{IEEEtran}
\usepackage{graphicx}
\usepackage{url}
\usepackage{booktabs}
\usepackage{subfigure}

\begin{document}

\title{Building Programmable Wireless Networks: \\\emph{An Architectural Survey}}
\author{Junaid Qadir, Nadeem Ahmed, Nauman Ahad}
\date{}
\maketitle

\begin{abstract}
In recent times, there have been a lot of efforts for improving the ossified Internet architecture in a bid to sustain unstinted growth and innovation. A major reason for the perceived architectural ossification is the lack of ability to program the network as a system. This situation has resulted partly from historical decisions in the original Internet design which emphasized decentralized network operations through co-located data and control planes on each network device. The situation for wireless networks is no different resulting in a lot of complexity and a plethora of largely incompatible wireless technologies.  The emergence of ``\emph{programmable wireless networks}'', that allow greater flexibility, ease of management and configurability, is a step in the right direction to overcome the aforementioned shortcomings of the wireless networks. In this paper, we provide a broad overview of the architectures proposed in literature for building programmable wireless networks focusing primarily on three popular techniques, i.e., software defined networks, cognitive radio networks, and virtualizable networks. In this paper, we provide a self-contained introduction on these techniques and its application and also discuss the opportunities and challenges in building next-generation programmable wireless networks.
\end{abstract}

\section{Introduction}

Wireless networks have become increasingly popular due to the inherent convenience of untethered communication. They are deployed ubiquitously in myriad of networking environments ranging from cellular mobile networking, regional or city wide networking (e.g., through WiMAX technology), local-area or even personal networking environments (e.g., through Wi-Fi and Bluetooth technology respectively) \cite{raychaudhuri2012frontiers}. With the usage of wireless networks promising to increase in the future, both in demand and application diversity \cite{akamai}, the issue of devising and implementing flexible architectural support becomes all the more important.

While newer wireless technologies have been emerging at a prolific rate, the architecture of wireless networking has largely been static and difficult to evolve. The malaise of architectural ``ossification'' is not unique to wireless networking though, but applies more generally to networking. Before we can describe the reasons of this ossification, we operationally define the \emph{data plane} to be responsible for forwarding packets at line speed, and the \emph{control plane} for figuring out, and instantiating, the forwarding state that the data plane needs. Various reasons have been offered to explain the Internet's architectural ossification such as: \emph{i)} vertical integration and coupling of the data plane and the control plane at node level, \emph{ii)} lack of abstractions and modularization of the control plane, and finally, resulting from the preceding two reasons: \emph{iii)} lack of programmability of the network as a whole. These reasons, subtly related to each other, have collectively discouraged networking growth and innovations \cite{koponen2011architecting}.

To manage the complexity of computer systems, computer scientists have long recognized the potency of the concept of abstraction \cite{liskov2010power}. It has been argued that the most formidable challenge to the networking industry is posed by the paucity of useful abstractions \cite{rexford2011networking}. With a lack of foundational abstractions, networking reduces to a ``plethora of protocols and tools'' without any underlying architectural base \cite{rexford2011networking} \cite{shenkerONS}. There are three main benefits of using \emph{abstractions}: \emph{i) modularity}, which allows managing complex problems scalably through reuse of modules offering common functionality, \emph{ii) separation of concerns} through loose inter-module coupling which is implementation agnostic, \emph{iii)  innovation}, since new developments can focus on the module that needs fixing without `reinventing the entire wheel' \cite{rexford2012report}.

The layered model is often considered as a major success story of computer networking, and the sustained scalability of the original Internet architecture decades after its commissioning is testament to this fact \cite{shenkerONS}. The famous OSI layering model, and the TCP/IP layering models, is composed of layers representing modular subcomponents that interact through well-defined abstract interfaces. While the data plane is layered, and offers some useful abstractions, the Internet's control plane has developed mostly in an ad-hoc fashion and has lacked well developed abstractions until quite recently \cite{mckeown2008openflow}. To develop \emph{programmable wireless networks}, it is imperative that we emphasize development of both \emph{programmable wireless data planes} and \emph{programmable wireless control planes} unlike existing schemes which have unfortunately focused on one or the other. In this paper, we will provide a unified holistic overview of programmable wireless networking and highlight overarching themes and insights.

Traditionally, the characteristics of the networking devices has offered certain `knobs', or \emph{configuration} options, that can be tuned to suit the operator.  The device's operator, however, is limited by the configuration options that are provided by the device's vendor. The vision of a programmable device is to allow the operator to \emph{program} in any way desired; i.e., the operator should be free to define new custom tunable knobs as desired to support niche applications or services.  Programmable devices thus offer far greater flexibility than configurable devices. This differentiation between \emph{configurability} and \emph{programmability} extends to network layer behavior as well. While traditional networks did provide limited configurability, the modern vision is to create fully programmable networks.

Broadly speaking, programmable networks denote networks that can tune itself, or reconfigure itself, through a software based interface. This software based adaptation is typically performed through an application programmer interface (API). Incorporation of functionality in software allows networks to innovate at the rate of software development cycle which is a lot more agile than the sluggish rate of hardware development \cite{andreessen2011software}. Applications of programmability include rapid provisioning of services \cite{Neutron}, flexible resource management \cite{mendoncca2013survey}, efficient resource sharing \cite{jain2013b4}, and support for new architectures such as cloud computing, Internet of things (IoT) \cite{ciscoIoE}, etc.

Interestingly, programmable networking is not entirely a recent concept. The lack of programmability of networks has long been recognized, and various approaches have been proposed to address this deficiency \cite{campbell1999survey} \cite{smith2004active}. In a remarkably prescient paper, published by Campbell et al. in 1999 \cite{campbell1999survey}, the impending impact of numerous programming trends is anticipated using surprisingly modern terminology. In particular, it was predicted that higher levels of network programmability will result from separation of hardware from software, availability of open network interfaces, virtualization of networking infrastructure, and rapid creation and deployment of network services. These predictions have come to fruition exactly as forecasted in the forms of software defined networking (SDN), standardized APIs, network virtualization, and cloud computing. Similarly, another insightful paper \cite{tennenhouse2002towards}, the early version of which dates to 1996, talks about applying programming language perspective to networks and their protocols, and talk about their aim of creating the `Smalltalk of networking'. These ideas are having a renaissance in the modern era in the context of SDN programming languages  \cite{foster2013languages}\footnote{Tennenhouse et al. also made the sobering observation in \cite{tennenhouse2002towards} that object-oriented approaches to networking are proposed every five to ten years with little impact on mainstream research.  Time will tell how transformative the modern proposals (e.g., SDN languages approach) will be in the long run.}.

There are three prominent technological trends that underlie most of the current research in future \emph{programmable wireless networking}. These promising technologies are:  \emph{i)} software defined wireless networks (SWN), \emph{ii)} cognitive wireless networks (CWN), \emph{iii)} Virtualizable Wireless networks (VWN). In this paper, we discuss these technological trends in detail, and provide both a self-contained tutorial as well as a detailed survey of the application of these trends.

The \emph{main contribution of this paper} is that we provide a unified overview of the emerging field of programmable wireless networks. We demonstrate that the seemingly disparate fields of active networking, software defined radios, cognitive radios, software defined networking, and wireless virtualization, are in fact kindred disciplines. We develop this idea and propose new directions of future programmable wireless networks. Our paper is different from other survey papers \cite{mendoncca2013survey} \cite{feamsterroad} \cite{campbell1999open} in its focus on \emph{wireless} programmable networks while the previous papers had focused mainly on generic (wired) programmable networks.

The organization of the remaining paper is as follows. In the next section (section \ref{sec:architecture}), we will describe the various architectural approaches to building programmable networks. Thereafter, we will follow it up with a detailed description of the various components of future programmable \emph{wireless} networks in section \ref{sec:progwireless}. We describe the three dominant categories of programmable wireless networks, i.e., SWNs, CWNs and VWNs, and highlight works belonging to each category in sections \ref{sec:SWN}, \ref{sec:CWN}, and \ref{sec:VWN} respectively. In section \ref{sec:openquestions}, we discuss various open research issues and future directions of research. We conclude the paper in section \ref{sec:conclusion}.

\section{Programmable Networking Architectures}
\label{sec:architecture}

With the lack of programmability complicating networking innovations, it was the early 1990s when work on creating programmable network started in earnest \cite{campbell1999open}. At the time there were two major, slightly differing schools, that advocated programmable networks: the first group proposed the `OpenSig' approach \cite{campbell1999open} while the second group furthered the `Active Networking' approach \cite{tennenhouse1997survey}. The general consensus that emerged was that the programmability solution lies in separating the control software from the hardware, and in having open interfaces for management and control. The building blocks for creating programmable networks started appearing thereafter in the form of various programmable networking components (such as the Click modular router \cite{morris1999click}, etc.). More recently, with the emergence of datacenters, virtualization, and cloud computing technology, the requirement of programmability has become mainstream. Many of the initial ideas of programmable networking (of `open interfaces' and `separation of control and data plane') espoused by the OpenSig and Active Networking community have now matured in the form of the `software defined networking' (SDN) architecture. Although SDN and active networking paradigms share a common motivation, i.e., of creating programmable networks, both of them are different in their focus: active networks strived more for data-plane programmability while SDN's focus has been on control-plane programmability \cite{feamsterroad}. In the remainder of this subsection, we will outline these developments in more detail.

\vspace{2mm}
\subsection{The OpenSig Approach}
\label{sec:opensig}

In the mid 1990s, the \emph{OpenSig approach} \cite{lazar1997programming} advocated both the separation of the data plane and the control plane for \emph{ATM networks} and the usage of \emph{open interfaces} for signalling between these two planes. The main idea was that with separated control and data planes, and an open standard interface, the ATM switches would become remotely programmable and thus more manageable. The OpenSig community actively worked on standardizing such an open interface, and a number of experimental networks set up in various research institutes explored these proposals. The Tempest framework \cite{rooney1998tempest}, based on the OpenSig philosophy, allowed multiple control planes to simultaneously control a single network of ATM switches. The main reason OpenSig approach could not quite become mainstream was the static nature of the interfaces it defined \cite{campbell1999survey}.

\vspace{2mm}
\subsection{Active Networking}
\label{sec:activeNetworking}

The \emph{Active Networking} (AN) approach \cite{tennenhouse2002towards}
\cite{tennenhouse1997survey} was popularized at the same time  as OpenSig, i.e., in the mid 1990s, when the Internet was rapidly commercializing and experiencing the need of more flexible control. The AN approach aimed at creating programmable networks that can allow rapid network innovations. The AN research---driven mainly by the efforts of the Defense Advanced Research Projects Agency (DARPA)---was motivated by the need to rapidly commission new services and dynamically configure networks in run-time. It was perceived that the static nature of OpenSig networks could not support these needs.

The main idea of AN is to \emph{actively} control network nodes so that the network nodes may be programmed to execute arbitrary mobile code as desired by the operator \cite{tennenhouse2002towards}. The value proposition of such an approach was that it would enable new innovative applications, that leverage computation with the network, and that it would increase the rate of innovations by decoupling services from the underlying infrastructure \cite{wetherall1998introducing} \cite{wetherall2002active}. The flexibility offered by such an approach, on the other hand, was also accompanied by concerns about its performance and security implications.

The AN approach consisted of two programming models: \emph{i)} the \emph{capsule model}, where the data packets contained not only the data to be communicated but also in-band instructions to execute, and the \emph{ii) programmable switch model}, in which the out-of-band mechanisms were utilized to execute code at various nodes \cite{feamsterroad} \cite{tennenhouse1997survey}. While it is the capsule model---which was the more radical approach, significantly different from the traditional operational paradigm of networking---that is most closely associated with AN, it is fair to say that both these models have bequeathed valuable legacies inherited by modern programmable networking frameworks.

In the \emph{capsule model}, special packets, or flows that consisted of actual program codes, were to be installed by controllers on smart nodes that ran a particular operating system (Node OS \cite{peterson1999nodeos}). The NodeOS project \cite{shalaby2002snow} focused on incorporating active networking technology into the Linux kernel while allowing regular non-active applications, and operating systems operations, to run unhindered without any significant performance penalty. The capsule approach attracted interest mostly since it could provide a clean method of upgrading data plane processing along an entire network path \cite{feamsterroad} \cite{wetherall2002active}. It was reported in \cite{wetherall2002active} that the most compelling application of capsules was actually enabling network layer service evolution and not necessarily the flexibility to run arbitrary code at any network location. Using the capsule approach, numerous services such as active load balancing, multicasting, caching etc. could be supported \cite{tennenhouse1997survey} \cite{psounis1999active}.

The AN framework was vigorously pursued by the research community in the mid and late 1990s---helped by the interest and generous funding of DARPA. Various influential projects were initiated in this time-frame with some prominent AN projects being the ActiveWare project \cite{activeware} at MIT, the CANES \cite{canes} project at Georgia Tech, the SwitchWare project \cite{switchware} at University of Pennsylvania, the ANTS project at University of Washington \cite{wetherall1999ants} \cite{ants}, and the Tempest project \cite{rooney1998tempest} at Cambridge University. More details about these, and other important AN projects, can be seen in table \ref{tab:concepts} and in the survey paper \cite{tennenhouse1997survey}.

Modern \emph{clean-slate} proposals such as SDN (which we will cover in section \ref{sec:sdn}) are indebted in large part to the AN community. The active networking paradigm was the first in a still continuing series of clean-slate Internet redesign proposals \cite{feamsterroad}. Many of the programmable networking concepts that appear eminently modern---such as separation of control plane and data plane, remote control of data planes, virtualization, network APIs, etc.---have in fact germinated from the active networking community.

It is now worthwhile to dissect the AN approach to highlight its deficiencies and to argue about its failure to capitalize on the intense interest around it. One reason for this failure is the lack of a compelling application use case in the AN approach that could work pragmatically within the existing framework. We shall see that while SDN architecture is very similar to the AN architecture, it appears to have become more mainstream due to technological advances, more compelling use cases, and importantly, certain pragmatic design choices. In particular, SDN has become popular largely due to the need of virtualization in modern datacenters and cloud computing which require network virtualization support due to their dependence on automated provisioning, automation, and orchestration. Another reason for AN's failure to become mainstream is its primary focus was on newer data plane abstractions while the SDN approach has focused more on newer control plane abstractions (which arguably addresses a bigger pain point). Thirdly, AN emphasized the flexibility of providing network end users the chance to program the network, which never became a popular use case, while SDN has focused more on wresting back control from network vendors and providing it to the network operator. Finally, the SDN architecture is different from the AN architecture since the former has emphasized on the separation of the control plane and the data plane \cite{feamsterroad} which was not integral to the AN architecture.

\vspace{2mm}
\subsection{Virtualization and Cloud Computing}
\label{sec:virtAndCC}

Virtualization is a technique, fundamental to various disciplines of computer science, which allows \emph{sharing} of resources while providing \emph{abstractions} identical, for all practical purposes, to that of the original resource.

Virtualization has been especially influential in the modern era of large-scale datacenters. Prior to the popularization of virtualization technology, various concerns (such as security, isolation, performance) dictated that servers be dedicated for particular applications (e.g., dedicated web servers, database servers, etc.) and provisioned for peak load. This led to gross under-utilization with 10\% -20\% utilization of resources being commonplace. This led to the creation of a new `virtual machine' (VM) abstraction using which multiple virtual machine instances, that were completely isolated from each other, would be created on the physical machine. These virtual machines provided an interface to end applications that was identical to that of the underlying physical server. With the programmability features of VM cloning and mobility, which allows taking VM snapshots and transporting to any physical server that is currently under-utilized, physical resources can now be shared both efficiently and securely. Due to these desirable properties, virtualization has truly become an indispensable component of modern computing.

The popularity of compute virtualization in the datacenter environment has spawned two further trends: \emph{i)} cloud computing, and, \emph{ii)} network virtualization (NV).

The main insight of ``\emph{cloud computing}'' is to provide services in a virtualized datacenter, provisioned programmatically through APIs via the web, as a service in \emph{utility computing} style. Although utility computing was conceived as early as 1961 by John McCarthy, it is only recently that cloud computing has turned this vision into a reality. The cloud paradigm is differentiated from traditional datacenters mainly in the dynamism of service provisioning which has been made feasible by virtualization technology and advances in web APIs. The ability to \emph{program} services has led to great innovations and has democratized computing largely by making computing resources available on per-use pricing. The `holy grail' of the cloud computing paradigm is the vision of installing a generic `network fabric' which can be then programmed to provide any service without any need of manual configuration of network nodes. The implementation of such a fabric based virtualized datacenter has proven itself elusive, due to the complexity of virtualizing networks, so much so that it is now a common sentiment in the networking industry that networking is the bottleneck in datacenter innovations\footnote{James Hamilton, the architect of Amazon's cloud, made the now famous remark in 2010 that ``datacenter networks are in my way'' bemoaning the lack of network programmability. See Hamilton's blog post at \url{http://perspectives.mvdirona.com/2010/10/31/DatacenterNetworksAreInMyWay.aspx} for more perspective.}. With traditionally vertically integrated network devices, supporting cloud-era applications entails the undesirable burden of manually configuring various network switches through vendor-specific command-line-interfaces (CLIs)---a process that is cumbersome and error prone \cite{networkdowntime}.

With the presence of 10s or 100s of VMs per machine, a software-based \emph{hypervisor switch}, inside the physical server, takes care of inter-VM networking. A significant tipping point was recently witnessed when estimated number of physical ports were overtaken by virtualized ports---a significant inflection point in networking history indeed \cite{crehan}. This has significant architectural implications. In particular, it has been highlighted that using an hypervisor overlay with a networking fabric constructed out of SDN technology (to be covered in section \ref{sec:sdn}) can become the functional equivalent of the traditionally influential \emph{end-to-end principle} \cite{shenkerStanford}. In addition, the virtualization/ SDN hybrid architecture will also subsume the functionality of MPLS and middleboxes to offer a clean split between the core and the edge. In this new architecture, the SDN based fabric will become the new core, while the hypervisor switches will be the new edge. We shall see later that these edge devices consist of hypervisor switches (e.g., Open vSwtich \cite{pfaff2009extending}) that are  \emph{software defined} and thus are \emph{programmable} (using protocols such as OpenFlow \cite{mckeown2008openflow}). This paradigm shift to software control fundamentally changes the pace of innovation, and opens up a world of new possibilities.

\begin{figure}[t]
\begin{center}
\includegraphics[width=.4\textwidth]{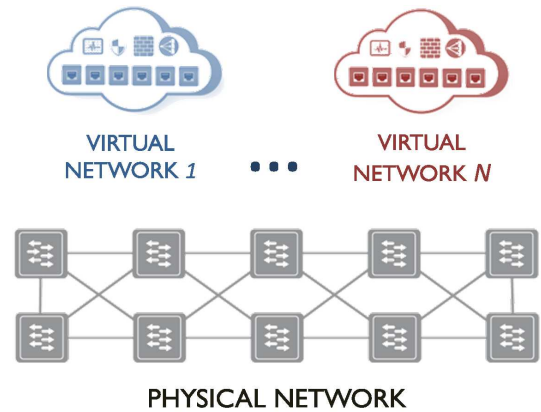}
\caption{Using virtualization, multiple virtual networks can coexist on the underlying physical infrastructure substrate in a decoupled fashion.}
\label{fig:vn}
\end{center}
\end{figure}

While VMs have unshackled applications from being tied to particular physical servers, traditional \emph{network virtualization} techniques (such as VLANs, VPNs, and overlay networks) do not offer an analogous ``\emph{virtual network}'' (VN) abstraction that decouples the network from the physical infrastructure. This VN abstraction should, like the VM abstraction does for the server, ensure detachment of the virtual network from the physical infrastructure as well as isolation between multiple tenants sharing the same infrastructure, while providing the same interface as the original network. There was a notable early work on network virtualization in the OpenSig era: the Genesis project \cite{campbell1999genesis} proposed, in 1999, a \emph{virtual network kernel} that was capable of \emph{spawning} virtual network architectures on-demand. The term `spawning' is used in Genesis as a metaphorical reference to the use of this term in the field of operating systems where it refers to the process of creating a new process that runs on the same hardware---analogously, spawning a network means creating a new network architecture on the same infrastructure. This concept, although important and novel, is distinct from the modern virtualization abstraction of a VN. Just like a VM is a software-container---encapsulating logical CPU, memory, storage, networking, etc.---providing an interface identical to a physical machine to an application, a VN is software container---encapsulating logical network components, such as routers, switches, firewalls, etc.---that presents an interface identical to a physical network to network applications. The VN abstraction for wireless networks is visually depicted in figure \ref{fig:vn}. This abstraction allows great flexibility to IT managers as the physical network can now be managed as a `fabric' offering some transport capacity that can be used, programmed, and repurposed as needed.

\begin{figure}[t]
\begin{center}
\includegraphics[width=.47\textwidth]{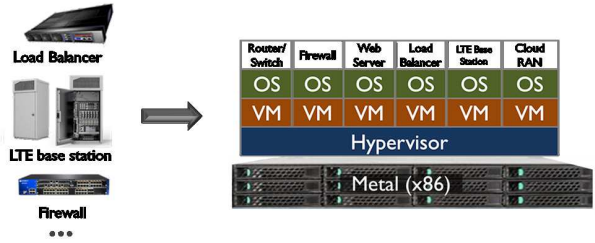}
\caption{Network functions virtualization (NFV) is used to convert fixed function hardware network appliances into virtualized cloud software instances that run on commodity infrastructure hosted in cloud datacenters.}
\label{fig:nfv}
\end{center}
\end{figure}

Virtualization is also a popular method in the Internet community for introducing innovations in production networks with minimal intervention through the use of \emph{overlay networks} \cite{anderson2005overcoming} \cite{chun2003planetlab}. There is also growing interest amongst major network service providers to decouple the functionality of telecom devices from dedicated devices to enable ``\emph{network functions virtualization}'' (NFV) (as can be seen in figure \ref{fig:nfv}) which will enable implementation of network functions (such as mobile network node, etc.) on servers in datacenters \cite{nfv}. With virtualization being the core functionality of almost all the recent important clean slate Internet redesign projects \cite{paul2011architectures} \cite{feldmann2007internet}, it is anticipated that virtualization will be the main technology, or the narrow waist, of future Internet's architecture. With its architectural promise and immediate commercial appeal \cite{nicira}, it is fair to say that virtualization has taken the cloud-era networking world by a storm.

The combination of cloud computing and network virtualization allow programmability that leads to unprecedented flexibility in rapidly creating, deploying, and managing novel services as per the demands of users. This can create a new service-oriented architecture for wireless networking where heterogeneous wireless access technologies may coexist and converge as \emph{extended cloud infrastructure} \cite{wen2013wirelessbook}. Thus the cloud computing concepts, proposed originally for datacenters, are likely to play a big part in creating future programmable \emph{wireless} networks.

Interested readers are referred to comprehensive surveys on network virtualization \cite{wang2013network} \cite{chowdhury2010survey} and cloud computing \cite{armbrust2010view} for more details.

\begin{table*}
\caption{Representative summary of important programmable networking concepts}
\centering
\begin{tabular}{p{2.6cm} p{1.9cm} p{.7cm} p{11.5cm}}
\toprule
\textbf{\emph{Project}} & \textbf{\emph{Framework}} & \textbf{\emph{Year}} & \textbf{\emph{Summary}} \\
\midrule

\\
\multicolumn{4}{l}{\textbf{\emph{Active Control of Network Nodes}}} \\
ANTS \cite{wetherall1999ants} \cite{ants} & Active Networking & 1997 & Java-based active networking toolkit proposed as part of the MIT's ActiveWare \cite{activeware} project \\
SwitchWare \cite{switchware} & Active Networking & 1998 & Active networking project at Uni. of Pennsylvania which focused on both security and performance issues \\
CANEs \cite{canes} & Active Networking & 1998/9 & Composable Active Network Elements (CANEs) project at Georgia Tech \\

\\\hline\\
\multicolumn{4}{l}{\textbf{\emph{Separation of the Control Plane (CP) and the Data Plane (DP), and Remote Control of DP}}} \\
Tempest \cite{rooney1998tempest}& OpenSig & 1998 & Programmable framework for safe control of ATM switches. Allowed multiple control architectures to coexist on the same network, and a safe partitioned environment for third party, or dynamically loaded, active code.
\\
GSMP \cite{doria2002general} & Pre-SDN & 2002 & General Switch Management Protocol (GSMP) proposed by an IETF working group to control a label switch\\
FoRCES \cite{yang2004rfc} & Pre-SDN& 2004 & It defines a standardized interface between a network's control elements (CEs) and forwarding elements (FEs)\\
RCP \cite{feamster2004case}& Pre-SDN& 2004 & This work proposed separating routing from routers, and outsourcing it to a separate router control platform (RCP)\\
SoftRouter \cite{lakshman2004softrouter} & Pre-SDN & 2004 & SoftRouter proposed separation of the control plane functions from the packet forwarding functions\\
4D \cite{greenberg2005clean}& Pre-SDN& 2005 & 4D proposed an architecture with \emph{decision}, \emph{dissemination}, \emph{discovery}, and \emph{data}---i.e., the 4D---planes, to separate decision logic from distributed systems issues\\
Routing as a service \cite{lakshminarayanan2004routing} & Pre-SDN & 2006 & Proposed the provision of offering customized route computation as a service by third-party providers\\
PCE architecture \cite{farrel2006path}& Pre-SDN & 2006 & Path-computation-element (PCE) based architecture (RFC 4655) where the PCE is an application located within a network node, or on an out-of-network server. \\
CogNet \cite{raychaudhuri2006cognet} & Pre-SDN & 2006 & CogNet proposed separated CP and DP with an extensible global CP controlling the separated DPs through an API\\

IRSCP \cite{van2006dynamic} & Pre-SDN & 2006 & Proposed Intelligent Route Service Control Point (IRSCP) that allowed route selection to be performed outside the routers through external network intelligence.\\

SANE \cite{casado2006sane}& Pre-SDN & 2006 & SANE is a  enterprise network security/ protection architecture  implemented through a ``logically-centralized'' server\\
Ethane \cite{casado2007ethane} & Cusp of SDN-era & 2007 & Ethane proposed fine-grain control of simple flow-based Ethernet switches through a centralized controller \\
OpenFlow \cite{mckeown2008openflow} & SDN & 2008 & OpenFlow defines a southbound API/ protocol standardized by ONF through which a separated dedicated controller can control multiple DPs remotely \\
\\
\hline
\\
\multicolumn{4}{l}{\textbf{\emph{Open APIs}}} \\
xbind \cite{lazar1996xbind}& OpenSig & 1996 & Toolkit developed at Columbia Uni. for creating \emph{broadband kernels}---that program broadband ATM nets like PCs\\
Mobiware\cite{angin1998mobiware}& OpenSig & 1998 & Programming QoS-aware middleware for mobile multimedia networking developed at Columbia University \\
NetScript & OpenSig & 1999& Language for programmable processing of packet streams\\
OpenFlow \cite{mckeown2008openflow}& SDN & 2008& Southbound API standardized by ONF  \\
Floodlight API \cite{mckeown2008openflow}& SDN & 2012& A RESTful northbound API between the controller platform and the SDN Applications \\
Juniper APIs \cite{juniper} & SDN & 2012& JunOS SDK, XML API (NetConf), and Contrail REST API supported \\
Cisco ONE \cite{onepk} & SDN & 2012& Network APIs (including southbound API) specified by Cisco; supports EEM (tcl), Python Scripting \\
OpenStack APIs \cite{openstack} & Cloud/ SDN & 2012& OpenStack Neutron (formerly Quantum) is a OpenStack subsystem for managing networks in a cloud environment \\

\\
\hline
\\
\multicolumn{4}{l}{\textbf{\emph{Network Virtualization}}} \\

Virtual Switches \cite{lazar1996xbind} & OpenSig & 1996 & Proposed virtualizing ATM switches as part of the xbind \cite{lazar1996xbind} project\\

Switchlets \cite{rooney1998tempest} & OpenSig & 1998 & Proposed dynamically loadable code on a (partition of) ATM switches as part of the Tempest \cite{rooney1998tempest} \\

Virtual base stations \cite{angin1998mobiware} & OpenSig & 1998 & Proposed as part of the Mobiware \cite{angin1998mobiware} project subscribing to the OpenSig framework\\

Routelets \cite{campbell1999genesis} & OpenSig & 1999 & Proposed in the Genesis \cite{campbell1999genesis} project\\

PlanetLab \cite{chun2003planetlab}\cite{anderson2005overcoming} & Overlays Networks &  2003 & Proposed overlays, virtualized slicing, and programmability for accelerating innovations in the Internet community. \\

FlowVisor \cite{sherwood2009flowvisor} & SDN & 2009 & Virtualizes OpenFlow based SDN environments by carving out virtualized ``\emph{slices}'' out of production networks \cite{sherwood2010carving}\\

SecondNet \cite{guo2010secondnet} & SDN/ Datacenters & 2012 & Proposes a virtual data center (VDC) abstraction as the unit of resource allocation for multiple tenants in the cloud\\
\\
\bottomrule
\end{tabular}
\label{tab:concepts}
\end{table*}

\vspace{2mm}
\subsection{Software Defined Networking (SDN)}
\label{sec:sdn}

SDN has revolutionized the networking industry by providing architectural support for ``programming the network''. SDN promises to be a major paradigm shift in networking landscape leading to improved and simplified networking management and operations. The major insight of SDNs is to allow horizontally integrated systems by allowing the separation of the control plane and the data plane \cite{mckeown2009software} \cite{mendoncca2013survey} while providing increasingly sophisticated set of abstractions. Although the term SDN has only been coined in 2009, the idea of SDN has a rich intellectual history. In particular, it is the culmination of many varied ideas and proposals in the general field of programmable networks \cite{mendoncca2013survey} \cite{feamsterroad}. While conceived mainly in academia, SDN has been taken up by the industry by gusto with numerous success stories \cite{nicira} \cite{contrail}. SDN has also been seen recent successful industrial deployments \cite{jain2013b4}.

\begin{figure}[!ht]
\centering
\subfigure[In traditional networking, the control planes (CP) and the data planes (DP) are co-located on devices to ensure decentralized network control.]{
 \includegraphics[width=.35\textwidth]{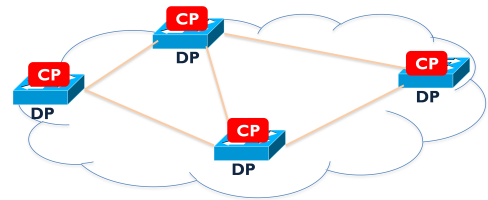}
   \label{fig:PreSDN}
 }
\subfigure[In SDNs, the DPs and CPs are separated with a centralized controller controlling multiple DPs while supporting a \emph{southbound} API to the DPs and a \emph{northbound} API to the SDN applications.]{
   \includegraphics[width=.45\textwidth]{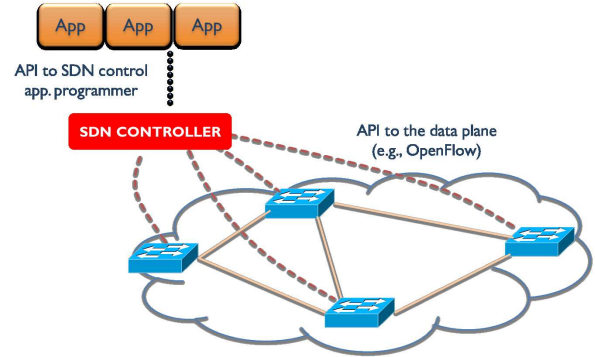}
   \label{fig:SDN}
}
\caption{Comparison of Traditional and SDN network architectures}
\label{fig:SDNoverall}
\end{figure}

With the growing popularity of SDN, various industrial stakeholders have jumped on the SDN bandwagon to exploit its early success, and the term `SDN' has seen a considerable broadening. To analyze and reason about SDN, it is, therefore, vital that we define it precisely. There are three key characteristics of SDN. Firstly, there is a separation of the data plane and the control plane. Secondly, a single control plane (or controller) may control multiple data planes (or the datapath of switches/ router). Lastly, SDNs incorporates modularity in the control plane through which high level abstractions can be used by network control programs. The distinction between traditional and SDN network architecture can be clearly observed in \ref{fig:SDN}. To summarize jointly both these views, SDN deals with abstractions and mechanisms for creating a general, horizontal networking platform.

By providing abstraction layers, it is possible to program new applications on central controllers for a wide variety of purposes. These \emph{applications} include setting up virtual networks, enforcing quality of service (QoS), explicit routing etc. The most fertile application area of SDN has been data centers and campus networks \cite{mckeown2008openflow}, however, SDN has also been proposed in many other settings such as: service providers, carrier networks, wireless networks. Specifically, SDN has been applied to wireless networks in varied settings such as wireless sensor networks (WSN), and wireless mesh networks (WMN) \cite{chung2012experiences}.

Although networks have always been software defined\footnote{David Clark, the former chair of the Internet Architecture Board (IAB), summarized the role of software implementations in the networking community ethos when he said, ``We reject kings, presidents and voting.  We believe in rough consensus and \emph{running code}''.}, writing, managing and updating the networking software could be done only by the developers employed by the vendors. This ``closed'' network architecture stifled innovation as the network was inherently non-programmable, and all new feature requests had to be routed to the networking vendor for implementation. SDN changes this paradigm by opening up the network through the simple, yet powerful, concept of separation of the control plane and the data plane. This separation, coupled with newer control abstractions, form the core of the new SDN architecture. The development of SDNs is supported by a burgeoning open-source community. With the separation of the control plane from the data plane, it is possible for third party/ open-source developers to write program applications for the controller. This allows networks to employ programmable commodity hardware rather than closed vendor hardware, increasing flexibility and development while reducing costs.

The Open Network Foundation (ONF) is an organization that is working on developing and maintaining standards for SDNs. Broadly speaking, there are two main application programmer interfaces (APIs) in the SDN architecture: \emph{i)} the \emph{Southbound API} defines an interface between a centralized network controller\footnote{The centralized SDN network controller can itself be built as a distributed system to be scalable and avoid a single point of failure.} and networking devices \cite{mckeown2008openflow}, while \emph{ii)} the \emph{Northbound API} defines the interface exposed by the controller to the network applications

\emph{OpenFlow} \cite{mckeown2008openflow} represents perhaps the most readily recognized protocol associated with SDN\footnote{OpenFlow should not be confused with the overall SDN \emph{architecture} since OpenFlow, popular as it is, is but only one \emph{protocol} that implements the southbound API as envisioned in the SDN architecture.}. OpenFlow is an example standard southbound API which has been standardized by the ONF. The standardization of OpenFlow has propelled it as the principal SDN control plane abstraction, enabling thereby numerous innovations \cite{laranetwork}. With the control logic implemented in a separate controller, and a standardized control API between the controller and the data planes, the vision of programming the network using a high-level control language can be achieved. While the current control API defined by OpenFlow is fairly primitive (and has been compared to Assembly language), it is a matter of time before higher level control languages are developed that offer more sophisticated abstractions. Indeed, work in this direction is already underway \cite{foster2013languages}. The seemingly innocuous refactoring of the functionality from individual devices to the centralized controller unleashes a powerful new paradigm offering abstractions which facilitate simplified, efficient, and scalable management of network operations and services.

OpenFlow has also been used for designing and prototyping high speed networking by a reusable platform \emph{OpenPipes} \cite{gibb2009openpipes}. Using an OpenFlow network, new systems can be constructed quickly by OpenPipes, like the Click modular router, by `plumbing' modules---be they implemented in CPU, FPGA, ASIC---together in a pipeline. OpenPipes also allows flexible migration of modules (implemented in software or hardware, or both) from one subsystem to another, even in a running system.

The development of the SDN architecture has also led to the development of the ``network operating system'' abstraction. The role of an operating system (OS) is to manage the complexity of various components, that a computer is composed of, and to present an simplified programming interface to the application programmer. In a similar vein, it is envisioned that the network OS (NOS) will manage the various tasks necessary to manage the network (such as exchange of distributed state and computation of routes, etc.) and present a simplified interface to network application programmers. The NOS is typically implemented at the SDN controller(s). An NOS is expected to implement a \emph{state management layer}, managing distributed state in the network to provide a consistent network view, and provide an API to network applications to facilitate high-level programming. Various NOS have been implemented for SDN including the seminal work for NOX \cite{gude2008nox}, and subsequent efforts for ONIX \cite{koponen2010onix} and ONOS \cite{ONOS}.

An initial SDN use case, espoused in \cite{mckeown2008openflow}, was allowing researchers to run experimental protocols in virtualized ``slices'' of the production network. The concept of slicing network through virtualization technology predates SDN, and has been used in the VINI \cite{bavier2006vini}, the PlanetLab \cite{chun2003planetlab} and the Emulab \cite{hibler2008large} projects, and more recently in the NSF funded Global Environment for Network Innovations (GENI) project \cite{elliott2010geni}. Taking this further, Casado et al. proposed the concept of a ``\emph{network hypervisor}'' to virtualize the network's forwarding plane \cite{casado2010virtualizing}. The term ``\emph{hypervisor}'' conventionally refers to a virtual machine monitor (VMM), which is a host program that runs on a physical machine, to control multiple virtualized machines (VMs) on that machine. The concept of a network hypervisor is analogous to the traditional hypervisor concept.  The network hypervisor implements a network-wide software layer through which it is aimed that multiple virtualized networks, that are decoupled from their underlying hardware instantiation, can be supported. In such an environment, the network state (forwarding and configuration) is decoupled from the underlying hardware, and thus networks can be created, moved, cloned, deleted just like VMs in the server world---all in software. A network hypervisor, FlowVisor \cite{sherwood2009flowvisor}, has been developed for OpenFlow based SDN environments that allows carving out of virtualized ``\emph{slices}'', that are isolated from each other and controlled by a separate NOS, out of OpenFlow production networks \cite{sherwood2010carving}.

The SDN architecture, as has been highlighted before, uses many of the programmability concepts of earlier projects. In particular, it builds upon earlier proposals to \emph{i)} separate the control plane and the data plane (early proposals include SoftRouter \cite{lakshman2004softrouter}, 4D \cite{greenberg2005clean}, RCP \cite{feamster2004case}, and work in the ForCES working group \cite{yang2004rfc}) \emph{ii)} control multiple data planes from a separate controller (like the Tempest framework \cite{rooney1998tempest}), \emph{iii)} utilize open interface for communications between the separate controller and the data planes (like the OpenSig framework \cite{lazar1997programming}). A representative summary of various programming concepts, many of which SDN exploits, is shown in table \ref{tab:concepts}.

Despite the fact that SDN utilizes many of the active networking projects, SDN has become more popular than its predecessors due to the various technology pushes (e.g., advances in computing and networking technology) and application pulls (e.g., datacenter and cloud services, network virtualization, etc.) and greater industrial acceptance due to certain pragmatic design choices \cite{feamsterroad}. The long-term success of SDN would requires innovations in new abstractions for the control and data plane balanced with a pragmatic strategy for its deployment.

It is pertinent here to clarify the connections between SDN and NV. Since both these technologies return some similar benefits, it is a common mistake to equate these two technologies \cite{feamsterroad}. The SDN architecture is characterized by its emphasis on the separation of the control plane and the data planes, and the potential management of multiple data planes through the separated control plane. NV, on the other hand, is characterized by its emphasis on a new 'virtual network' (VN) abstraction that decouples the virtual network from the physical infrastructure. It is another myth that NV is just an application of SDN. It is worth stressing that NV is a \emph{solution} while SDN is an \emph{architecture}---while NV can be implemented more easily using the architectural flexibility offered by SDN, implementation of SDN is not a prerequisite for NV. It has been argued quite convincingly that NV is a distinct entity, important in its own right \cite{brucedavie} \cite{feamsterroad} \cite{shenkerStanford}, which may turn out to be even bigger than the current SDN fad sweeping the networking community  \cite{shenkerStanford}.

The development of \emph{programmable wireless networks}, as highlighted before, requires changes not only in the control plane but also in the data plane. In particular, the wireless data plane needs to be redesigned to define new, more useful, abstractions. To put things into perspective, the current data plane abstraction offered by OpenFlow supporting switches is based on primitive ``\emph{match-action}'' paradigm. To lend greater support to innovations in control plane functionality, the data plane functionality has to evolve to support more sophisticated, and useful, abstractions.  Research on newer data plane abstractions is being vigorously pursued with the use of programmable hardware being popularly proposed \cite{anwer2010building}. To reiterate the central thesis of this paper, the vision of programmable wireless networks requires synergy in multiple related domains and would require innovations in both the data plane and the control plane of wireless networks.

Since the proposal of the SDN architecture \cite{mckeown2008openflow} in 2008, many research works have focused on the development of higher layer protocols and applications that can leverage and exploit the programmability offered by the SDN architecture. In particular, routing, transport-layer, and management frameworks have been proposed that work with OpenFlow and SDN. The routing proposals include \emph{i)} Quagflow \cite{nascimento2010quagflow}, which partners the open-source routing software Quaqqa \cite{quagga} with OpenFlow , \emph{ii)} RouteFlow approach \cite{nascimentorouteflow} which can be used to provide `virtual routers as a service' in SDN environments, \cite{nascimento2011virtual}. The transport layer protocol proposals include the work on OpenTCP \cite{ghobadi2012rethinking}. Finally, there has been work on supporting multimedia delivery with QoS with the OpenQoS \cite{egilmez2012openqos}.

\vspace{2mm}
\section{Building Blocks for Programmable Wireless Networking}
\label{sec:progwireless}

Programmable devices are envisioned to be a key component of future programmable networks. In this section, we discuss various techniques and architectures that have been proposed to realize the benefits associated with programmable wireless networks. In particular, we elaborate upon the trends of software defined radio, cognitive radio, MAC programmable wireless devices, programmable wireless testbeds, and programmable radios in this particular order.

\vspace{2mm}
\subsection{Software Defined Radios (SDR)}
\label{sec:sdr}

The defining characteristic of a SDR is that it implements most of the basic building blocks of PHY layer radio communication in software. With the hardware stripped down to the elements essential to all radio communication, custom blocks that were implemented in hardware traditionally---e.g., filters, amplifiers, modulators, demodulators, mixers, etc.---are now implemented in software. This implies that appropriate programming of the generic radio hardware can in principle allow it to support arbitrary technologies. The SDR technology was a significant paradigm shift ushering in a new era of programmable wireless devices. Thus, using an SDR, an operator could program a wireless device to support any of the myriad of wireless technologies \cite{burgess2008openbts}. This opened up an unprecedented opportunity for creating a programmable \emph{wireless} device for the first time \cite{reed2002software} \cite{tuttlebee2003software} \cite{ulversoy2010software}.

The precise definition of SDRs is debated, with no clear consensus on how reconfigurable must a radio be to be deemed an SDR. Clearly, it is a bit of a stretch to call every radio with a digital signal processor (DSP) as an SDR. A working definition provided in \cite{reed2002software} is that an SDR is ``a radio that is substantially defined in software and whose physical layer behavior can be significantly altered through changes to its software''.  The SDR forum defines a `ultimate software radio' (USR) as ``a radio that accepts fully programmable traffic and control information and supports a broad range of frequencies, air-interfaces, and applications software.'' \cite{reed2002software}.  In \cite{partridge2011realizing}, two extremes SDR platforms are discussed: The first type is an SDR  that is composed of programmable components, such as field programmable gate arrays (FPGAs), DSPs, etc., which are \emph{programmed} directly; the other extreme is a highly \emph{configurable} chipset based SDR which is `programmed' by setting configuration registers in the chip to determine the choice of frequency, coding, and PHY and MAC level protocol details. Most practical SDRs lie between these two extremes \cite{partridge2011realizing}.

While traditionally SDRs have mainly been used in military settings due to excessive cost, the technology has now matured to a stage where its form and cost is amenable to non-military markets \cite{partridge2011realizing}. While the SDR of 1990s was the size of a small refrigerator and could easily cost more than \$100,000, today the size of an SDR is akin to the size of a compute battery and it can cost less than \$500, extrapolating the trend, it is reasonable to assume that future pricing and form factor of SDRs will match that of a typical consumer electronic device \cite{partridge2011realizing}. The democratization of SDR technology will conceivably revolutionize wireless and mobile networking: e.g., a consumer will not be limited to any single wireless protocol with a wireless device. This will lead to unprecedented flexibility as technologies (such as Wi-Fi and Bluetooth) will no longer be `baked' into the hardware, but will be software applications, or applets, that any SDR could support. Due to their versatile nature, SDRs are radio chameleons potentially running a telephony protocol (such as CDMA) at a given time, and switching to a completely different data communication protocol (such as Wi-Fi or WiMAX) next moment \cite{partridge2011realizing}.

A prominent, and popular, example of SDR platform is to use the universal software radio peripheral (USRP) hardware kit \cite{ettus} along with the open-source GNU Radio software toolkit \cite{gnuradio} that implements in software various necessary signal processing blocks. The USRP hardware digitizes the received analog signal, and imports  it into a computer so that it may be processed by GNU radio software (or similar toolkits such as the OSSIE framework based on the JTRS Software Communications Architecture \cite{ossie}). Such an arrangement allows building a \emph{custom radio} that can be programmed to support an arbitrary wireless technology through appropriate signal processing blocks that operate on the signal received, or to be transmitted. Other SDR examples include WARP \cite{khattab2008warp}, SORA \cite{tan2011sora}, and Airblue \cite{ng2010airblue}.

SDRs are envisioned to be a essential component of future programmable wireless devices. In particular, their importance can be gauged from the fact that almost all advanced wireless programmability techniques (such as cognitive radio, and programmable wireless processors, etc.) are based on SDRs.

\begin{table*}
\caption{Representative summary of important programmable networking components and platforms}
\centering
\begin{tabular}{p{3.5cm} p{3.3cm} p{0.4cm} p{9.3cm}}
\toprule
\textbf{\emph{Component Category}} & \textbf{\emph{Project and Reference}} & \textbf{\emph{Year}} & \textbf{\emph{Brief Summary}} \\
\midrule

\multicolumn{3}{l}{\textbf{\emph{Software defined radio (SDR) platforms}}} \\
& IRIS \cite{mackenzie2004software} & 2004 & Implementing Radio in Software (IRIS) project developed at Trinity College, Dublin\\
& USRP \cite{ettus2005usrp} & 2005 & Flexible SDR development platform, often used with GNUradio, manufactured by Ettus/ NI \\
& WARP \cite{khattab2008warp} & 2008 & Wireless Open-Access Research Platform (WARP) developed by Rice University\\
& SORA \cite{tan2011sora} & 2011 & Programmable SDR platform, developed by Microsoft, for commodity multi-core PCs\\
& OpenRadio \cite{bansal2012openradio} & 2012& Programmable wireless dataplane that can programmed across the wireless stack\\
\\
\hline
\\
\multicolumn{3}{l}{\textbf{\emph{Cognitive radio (CR) platforms}}} \\
& BEE2 \cite{mishra2005real} & 2005 & Reconfigurable hardware platform developed at University of California, Berkeley\\
& KNOWS \cite{yuan2007knows}& 2007 & CR hardware platform, for operation in TV whitespaces, developed by Microsoft\\
& WinC2R \cite{miljanic2008winlab} & 2008 & CR hardware platform developed by the WINLAB at Rutgers University\\
\\
\hline
\\
\multicolumn{3}{l}{\textbf{\emph{Programmable network components}}} \\
& Virtual Switches \cite{lazar1996xbind} & 1996 & Proposed virtualizing ATM switches as part of the xbind \cite{lazar1996xbind} project (OpenSig framework) \\

& Switchlets \cite{rooney1998tempest} & 1998 & Proposed dynamically loadable code on a (partition of) ATM switches as part of the Tempest \cite{rooney1998tempest} project subscribing to OpenSig framework \\

& Virtual base stations \cite{angin1998mobiware} & 1998 & Proposed as part of the Mobiware \cite{angin1998mobiware} project subscribing to the OpenSig framework\\

& Routelets \cite{campbell1999genesis} & 1999 & proposed in the Genesis \cite{campbell1999genesis} project subscribing to the OpenSig framework\\

& Click \cite{morris1999click} & 1999 & Software architecture for building flexible and configurable routers \\
& XORP \cite{handley2003xorp} & 2003 & An open programmable router platform for research experimentation \\

& NetFPGA \cite{lockwood2007netfpga} \cite{naous2008netfpga} & 2007 & Programmable and extensible router with embedded FPGA  \\

& RouteBricks \cite{dobrescu2009routebricks} & 2009 & Software router architecture (Click based) that parallelizes router functionality\\

& SwitchBlade \cite{anwer2010switchblade} & 2010 & FPGA based platform for deploying custom protocols with programmability and performance\\

& Ansari et al. \cite{ansari2010decomposable} & 2010 & Programmable decomposable MAC framework \\

& TRUMP \cite{zhang2011trump} & 2011 & Programmable component-based MAC framework\\

& Wireless MAC processor \cite{tinnirello2012wireless} & 2012& Composition of custom MAC protocols by programming with basic MAC commands\\

& MAClets approach \cite{bianchi2012maclets} & 2012 & Programmable framework that allows installing MAC stacks as `applets'\\

\\ \bottomrule
\end{tabular}
\label{tab:components}
\end{table*}

\vspace{2mm}
\subsection{Cognitive Radios (CR)}
\label{sec:cr}

CRs have evolved from the concept of SDRs \cite{jondral2005software}.  Joseph Mitola coined the term ``cognitive radio'' in 1999 when he envisioned a broadening of the SDR concept. In particular, Mitola anticipated that incorporation of substantial artificial intelligence (AI), in the form of machine learning, knowledge reasoning and natural language processing will transform SDRs into intelligent radios that can sense, learn, and react to network conditions to satisfy some notion of optimality \cite{mitola2006cognitive}. In a modern setting, this is achieved by incorporation of a cognitive engine that employs various AI-based techniques to build a knowledge base, based on which reasoning is performed to make ‘optimal’ decisions \cite{he2010survey}. In a nutshell, CRs evolves from the SDR concept, and allows an SDR to reprogram itself autonomously based on network conditions. After SDR technology, CRs represented the next big shift in the drive towards powerful programmable wireless \emph{devices}.

CRs are viewed as an essential component of next-generation wireless networks \cite{akyildiz2006next} \cite{haykin2005cognitive}, and have a wide range of applications including intelligent transport systems, public safety systems, femtocells, cooperative networks, dynamic spectrum access, and smart grid communications \cite{akyildiz2006next} \cite{he2010survey}. CR can dramatically improve spectrum access, capacity, and link performance while also incorporating the needs and the context of the user \cite{he2010survey}.  Although cognitive behavior of CRs can enable diverse applications, perhaps the most cited application of cognitive radio networks (CRNs), which are networks where nodes are equipped with CRs, is dynamic spectrum access (DSA) \cite{fette2009cognitive}. DSA is proposed as a solution to the problem of ‘artificial spectrum scarcity’ that results from static allocation of available wireless spectrum using the command-and-control licensing approach \cite{fette2009cognitive}. Under this approach, licensed applications represented by primary users (PUs) are allocated exclusive access to portions of the available wireless spectrum prohibiting other users from access even when the spectrum is idle. With most of the radio spectrum already being licensed in this fashion, innovation in wireless technology is constrained. The problem is compounded by the observation, replicated in numerous measurement based studies world over, that the licensed spectrum is grossly underutilized \cite{akyildiz2006next} \cite{fette2009cognitive}. The DSA paradigm proposes to allow secondary users (SUs) access to the licensed spectrum subject to the condition that SUs do not interfere with the operations of the primary network of incumbents.

While programmable wireless devices (such as SDRs and CRs) do serve as the building block for programmable wireless networking infrastructures, it is pertinent to note here that the task of building programmable wireless \emph{networks} is much more nuanced. Various vexing problems (such as routing \cite{qadir2013artificial}, security \cite{baldini2012security}, etc.) need to be solved while taking into account network wide behavior \cite{raychaudhuri2006cognet}. Historically, most of the CR research has focused on optimizing at a device level, with network level programmability being a recent afterthought \cite{thomas2007cognitive}. In subsequent sections, we will see how trends of cognitive networks (section \ref{sec:CWN}) and software defined networks (section \ref{sec:SWN}) allow us to extend the programmability concepts to network proportions.

\vspace{2mm}
\subsection{MAC Programmable Wireless Devices}
\label{sec:progwirelessproc}

In the past few years, numerous new wireless technologies, with distinct MAC protocols, have been proposed to serve a variety of niche wireless applications. Since there is no universal, one-size-fits-all, MAC protocol that will work equally well for all such scenarios, there is a lot of interest in creating programmable wireless devices which will implement, what may be effectively called, \emph{software defined MAC}. A majority of current wireless devices do not support SDRs, or even software defined MAC, and effectively can support only a single technology. Although SDRs offer great flexibility in altering its PHY later characteristics, supporting programmatic MAC on SDRs also entails significant research challenges \cite{nychis2009enabling} \cite{tan2011sora}.

In recent times, there has been work in supporting programmable, or software defined MAC, on \emph{commodity wireless devices}. One way of doing this is by creating an abstraction of a ``\emph{wireless MAC processor}'' with an instruction set representing common MAC actions, events and conditions which can be programmed through an API to compose any custom MAC protocol \cite{tinnirello2012wireless}. Another approach, known as the MAClet approach, is to conceive the entire MAC protocol stack akin to a Java applet which can be loaded onto a MAC processor and run \cite{bianchi2012maclets}. While these approaches could be conceivably implemented on FPGA based SDR platforms, such as WARP \cite{khattab2008warp} or USRP \cite{ettus} in a straight-forward manner, the main contribution of the works \cite{tinnirello2012wireless} \cite{bianchi2012maclets} has been to implement these approaches on a commodity Broadcom wireless NIC. In \cite{gallo2012breaking}, a new service-oriented architecture for programmable wireless interfaces is proposed which replaces the traditional PHY and MAC layers with a \emph{i) a platform layer}, which exposes static primitives for managing hardware events and frame transmissions, \emph{ii) three layers of functionalities}---state machines, functions, and services---that expose a programmable interface to upper layers.  The proposed approach differs from SDR solutions since the adaptation and customization is  accomplished through programmable interfaces exposed at a layer higher than the PHY layer. Besides these aforementioned works \cite{tinnirello2012wireless} \cite{bianchi2012maclets} \cite{gallo2012breaking}, there have been other ``\emph{component oriented design}'' \cite{messerschmitt2007rethinking} efforts for composing customizable MAC protocols from a set of basic functional components \cite{ansari2010decomposable} \cite{zhang2011trump}.

A representative summary of various architectural components of programmable networking, including a summary of programmable MAC devices, is provided in table \ref{tab:components}. For a detailed survey of dynamically adaptable protocol stacks in general, the interested reader is referred to \cite{gazis2010survey}.

\vspace{2mm}
\subsection{Programmable Routers}
\label{sec:progrouters}

Programmable routers have been developed that incorporate programmable data path processing capabilities to perform custom protocol operations and/ or any arbitrary payload processing. These programmable routers are not specific to wireless technologies but we discuss this technology in this section because these routers can potentially be very useful in the context of programmable wireless networking. The Click programmable router \cite{morris1999click} is an early influential software router which snaps together various modular `elements' to assemble the router logic. Although Click offers the capability of rapid prototyping and deployment and decent performance for a software router running on a PC, any purely software-based approach will be hard pressed to satisfy the demanding performance requirements of modern networks. More recently, programmable routers with reprogrammable hardware such as FPGAs have been proposed to simultaneously address the needs for flexibility, extensibility, and performance for the forwarding-plane. Prominent projects in this category include the NetFPGA project \cite{lockwood2007netfpga}, the RouteBricks project \cite{dobrescu2009routebricks}, and the SwitchBlade project \cite{anwer2010switchblade}. Extensible open-source \emph{control plane software} also exists with the XORP open source software suite \cite{handley2003xorp} being a prominent example; XORP defines a fully extensible platform, suitable for both research and deployment, which builds upon the extensible Click framework in its forwarding plane.

\section{Three Dominant Trends in Programmable Wireless Networking}
\label{sec:threetrends}

In this section, we focus on three prominent trends in wireless networking that have the potential to play a major part in creating future programmable wireless networks. In particular, we discuss software defined wireless networks (SWNs), cognitive wireless networks (CWNs), and virtualizable wireless networks (VWNs) in sections \ref{sec:SWN}, \ref{sec:CWN}, \ref{sec:VWN}, respectively. Our generalized treatment of wireless networking will subsume discussions on both technologies that have evolved from their telecom roots (such as 4G networks such as WiMAX and LTE) and also those that have pre-dominantly data networking foundations (such as Wi-Fi).

\begin{table*}
\caption{Representative summary of important trends in wireless networking}
\centering
\begin{tabular}{p{2.3cm} p{5cm}  p{9.2cm} }
\toprule
\textbf{\emph{Application}} & \textbf{\emph{References}} & \textbf{\emph{Main Idea(s)}}  \\
\midrule
\\
\multicolumn{2}{l}{\textbf{\emph{Trend 1: Software Defined Wireless Networks (SWNs)}}}\\
\\
WLAN-based SWNs & Odin \cite{suresh2012towards}& SDN benefits include flexible control, better management, rapid innovations, etc. \\

Cellular Mobile SWNs & MobileFlow \cite{pentikousis2013mobileflow}, SoftRAN \cite{gudipati2013softran}, SoftCell \cite{jinsoftcell}&  Benefits include better radio resource management, real-time monitoring, flexible routing, better mobility support, and the ability to offload data to Wi-Fi networks \\

WSN-based SWNs & Luo et al. \cite{luo2012sensor}&  Using SDN principles in WSNs allow the usual SDN benefits (flexibility, rapid innovation, optimized resource utilization, etc.) \\

LRPAN-based SWNs & Costanzo et al.\cite{costanzo2012software}& Benefits include simpler management, flexible control, and more efficient resource utilization \\
\\
\hline
\\
\multicolumn{2}{l}{\textbf{\emph{Trend 2: Cognitive Wireless Networks (CWNs)}}}\\
\\
DSA-based CWNs & DSA Survey \cite{zhao2007survey}& Allows a secondary network to coexist with incumbent users belonging to the primary net. \\
Cloud-based CWNs & TV-white-space and clouds \cite{ko2011cooperative}& CWNs can perform increasingly complex tasks by offloading these computations to the cloud \\
\\
\hline
\\
\multicolumn{2}{l}{\textbf{\emph{Trend 3: Virtualizable Wireless Networks (VWNs)}}}\\
\\
WLAN-based VWNs & Commodity WLAN card VWN \cite{smith2007wireless},Virtual APs \cite{hamaguchi2010framework}, Virtual Wi-Fi \cite{xia2011virtual}, MPAP \cite{he2010mpap}&  Virtualization allows better support for multi-tenancy and multi-provider and infrastructure sharing, which is convenient both in terms of user experience and economics\\

Cellular Mobile VWNs & RAN Virtualization \cite{costa2013radio}, WiMAX BS virtualization \cite{bhanage2010virtual}, LTE virtualization \cite{zaki2010lte} & Support for multi-tenancy, infrastructure sharing, multiple virual network operators (MVNOs), infrastructural sharing etc.\\

Cloud-based VWNs & CloudMAC \cite{vestin2013cloudmac}, Wireless net. as a service \cite{vassilaras2011wireless}, Wireless network cloud \cite{lin2010wireless}. & VWNs can benefit from the cost and scalability benefits of cloud computing. The important new paradigm of mobile cloud computing \cite{dinh2011survey} opens up many possibilities\\
CRN-based VWNs & Spectrum Virtualization Layer \cite{tan2012enable} & This work proposed  a virtualized layer for supporting dynamic spectrum access in general wireless networks \\
\\
 \bottomrule
\end{tabular}
\label{tab:trends}
\end{table*}

\subsection{Trend 1: Software Defined Wireless Networks (SWNs)}
\label{sec:SWN}

With increasing deployment, and diversification of wireless technology, managing wireless networks has become very challenging. SDN is a promising architecture that can be used  for conveniently operating, controlling, and managing wireless networks. As discussed in section \ref{sec:sdn}, the defining characteristic of SDN is generally understood to be the separation of the control and data plane. The presence of programmable controllers enables us to call these networks `software defined'. Using SDN technology  for wireless networking will extend the benefits of SDNs---simplification, flexibility, evolvability, and rapid innovation---to wireless environments \cite{costanzo2012software}.

In the remainder of this section, we detail different wireless networking projects that have incorporated SDN principles. These projects vary in the manner in which they employ SDN principles as well as in the nature of wireless networks (sensor networks, cellular networks, etc.).

\vspace{2mm}
\subsubsection{OpenRadio}

The ``\emph{OpenRadio}'' system \cite{bansal2012openradio} defines a novel design of a wireless dataplane that allows programming of the entire wireless stack through a modular and declarative programming interface. OpenRadio proposes to refactor the functionality of wireless protocols into two parts. The \emph{processing plane} deals with programs and algorithms that process data using the underlying hardware. The \emph{decision plane}, on the other hand, is responsible for making logical decisions on the data being processed by the processing planes. It should be observed that the concepts of the processing and decision planes are subtly analogous to that of the data and control planes in the SDN world, respectively.

OpenRadio is themed in the mold of both SDRs and SDNs. OpenRadio uses an abstraction layer for managing wireless protocols on generic hardware configured through software like SDRs, while also allowing the separation of protocol from hardware similar to SDNs. OpenRadio can support different wireless protocols, like Wi-Fi, WiMAX and LTE, etc. though a common hardware, thereby significantly reducing costs and making it easier to configure, optimize and even define protocols. OpenRadio's major strength is its ability to detach protocol from hardware and to bind the former with software to allow increased flexibility. With newer wireless protocols regularly being rolled out, the ability to reprogram functionality centrally and programmatically is of great convenience. OpenRadio can also be used for cell-size based optimization in cellular networks and for management of frequency spectrum in the presence of multiple heterogeneous cell stations to avoid interference \cite{bansal2012openradio}.

\vspace{2mm}
\subsubsection{OpenRoads or `OpenFlow Wireless'}

A seminal development in the field of programmable SWNs has been the ``\emph{OpenRoads}'' project \cite{yap2010openroads}---known also as \emph{OpenFlow Wireless} \cite{yap2010blueprint}. OpenRoads provides a complete platform that can be used to apply SDN principles in wireless environments, and thereby create a programmable wireless data plane. One particularly appealing benefit of OpenRoads is that it allows efficient handover between diverse wireless technologies, by `flattening' multiple vertically integrated wireless technologies, to allow seamless mobility for clients of mobile wireless networks. In \cite{yap2010blueprint}, the feasibility of this SDN-based approach is explored for mobility management with vertical handovers between IEEE 802.11 and IEEE 802.16 networks. OpenRoads employs OpenFlow and the Simple Network Management Protocol (SNMP) on wireless routers. OpenFlow provides means to manage the forwarding plane while SNMP allows configuration of these wireless devices. Flow-visor and SNMP demultiplexer are used to divide and make the control more scalable. Each controlling flow is given a particular `flow value' to ensure that different controlling flows are isolated from one another so that only those flows that are intended for particular devices would be installed. High level control interfaces are used upon OpenFlow to provide communication between different devices, configuration of these wireless devices and management of flow.

\vspace{2mm}
\subsubsection{WLAN-based SWNs}

\emph{Odin} \cite{suresh2012towards} is a proposed SWN architecture that employs the principles of SDNs in wireless local area networks (WLANs). In its popular form, WLAN decisions are made by clients and not the WLAN infrastructure itself. For example, a client decides which access point it prefers to join rather than the  infrastructure deciding it for the client. In WLANs, association of clients with specific access points keeps on changing with client mobility. This poses a significant challenge to any potential SDN oriented WLAN architecture as it would be difficult for controller programmers to keep track of the ever changing association between access points and clients. The Odin architecture suggests the usage of light virtual access points (LVAPs). LVAPs virtualize access point-client association and decouples it from physical access points. Whenever a client connects to the WLAN network, it is allotted an identification number on its LVAP that remains fixed regardless of its associated physical access point. The complexities of the physical access point are thus hidden from central controller programmers. The Odin program offers many \emph{advantages}. Odin provides seamless mobility between access points as the need to constantly establish new connections with physical access points changes. Additionally, flexible routing policies further allow load balancing. Furthermore, with an improved overview of the network, it is possible to reduce interference and eliminate issues such as hidden node problems.

\vspace{2mm}
\subsubsection{Cellular Mobile SWNs}

Recently, there has been significant interest in improving the performance of cellular mobile networks through SDN principles. In particular, frameworks have been proposed that incorporate SDN principles into 3GPP evolved packet core (EPC) mobile carrier networks (the MobileFlow project \cite{pentikousis2013mobileflow}) and 4G long term evolution (LTE) cellular networks (the SoftRAN project \cite{gudipati2013softran} and the SoftCell project \cite{jinsoftcell}). The main \emph{advantages} of the cellular mobile SWN approach include better management of radio resources, more flexible routing, real-time monitoring, better mobility support, and the ability to offload data to Wi-Fi networks \cite{li2012toward} \cite{OffloadWiFi}.

There are a few problems with the current LTE architecture: \emph{i)} centralized data flow as all the data passes through the packet gateway (P-GW), \emph{ii)} centralized monitoring and control is not scalable and is expensive., \emph{iii)} base stations and infrastructure are difficult to configure. The first two problems are related with the central control and monitoring of LTE networks. Thus a possible solution would need to distribute some of the control and monitoring responsibilities leading to a hybridized control plane. This seems to be a departure from one of the fundamental principles of SDNs, i.e., centralized control.  Solution to the third problem lies in adapting an SDN based architecture so that remote applications may be used for the tasks. As discussed earlier, OpenRadio provide an ideal modular interface to configure base stations remotely and conveniently.

\vspace{2mm}
\subsubsection{Wireless Sensor Network (WSN)-based SWNs}

Wireless sensor networks (WSNs) have been popular within the research community but have always being considered as an application specific technology. Treating WSNs as application specific technology leads to the problem of resource under-utilization with potentially multiple application-specific WSNs being deployed over a shared area where a single WSN may have sufficed. Incorporating SDN in sensor networks would provide solution to these problems \cite{luo2012sensor}. Separation of control and data planes would provide abstraction, helping to manage and control the network. By employing sensor-network based SWN, network controllers could set policies and quality of services to support multiple potential applications. This would also allow usage of the same sensor nodes for multiple application/ purposes. This again increases resource utilization and optimization.

\vspace{2mm}
\subsubsection{Low Rate Personal Area Network (LR-PAN)-based SWNs}

SDN attributes can also be used to great advantage in LR-PANs \cite{costanzo2012software}. All LR-PANs essentially employ the same 802.15 data link layer \cite{zheng2004comprehensive}. Differences in higher layers of their respective protocol stacks lead to different LR-PAN protocols such as ZigBee, Bluetooth etc. This leads to incompatibilities in communication between different nodes. By using the same tools that are used in OpenRadio, we could separate hardware from protocol and use an abstraction layer to program and define different wireless protocols. This would allow us to run different wireless LR-PAN protocols on the same wireless device. In this way it would be possible for nodes to be dynamically associated with many networks at a time, allowing us to use network resources more efficiently. The separation of data and control planes extend the usual SDN benefits of simpler management, flexible control, and efficient resource utilization to LR-PAN SWNs.

\vspace{2mm}
\subsubsection{General comments about SWNs}

By analyzing various projects on SWNs, it has been observed that while the finer details are application dependent, all SWNs seek to: \emph{i)} attempt to make management of networks a lot more easier, \emph{ii)} allow the same hardware to support multiple wireless protocols, and \emph{iii)} provide an abstraction layer to allow all, or some part, of the wireless architecture to be programmable. These aims are facilitated through the separation of the control and data planes which allows a separate controller to programmatically reconfigure network properties. The theme of providing abstractions for programmability thus pervades the SWN approaches we have discussed in this section.

\subsection{Trend 2: Cognitive Wireless Networks (CWNs)}
\label{sec:CWN}

\begin{figure}[t]
\begin{center}
\includegraphics[width=.4\textwidth]{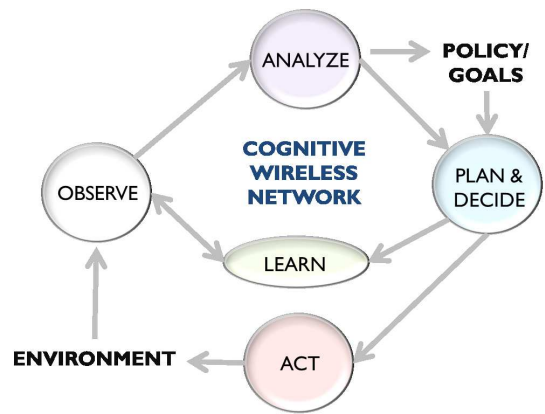}
\caption{Cognitive wireless networks (CWN) include an embedded cognitive engine which can observe network conditions, orient itself with the context, learn from experience, and decide to act, all while taking into account end-to-end network goals.}
\label{fig:ooda}
\end{center}
\end{figure}

It has been highlighted earlier that the predominant focus of most of the existing CRN research has been on enabling intelligent device-level behavior, with a notable exception being some work on \emph{cognitive networks} \cite{raychaudhuri2006cognet} \cite{thomas2007cognitive} \cite{mahmoud2007cognitive} \cite{manoj2007architectures} \cite{fortuna2009trends}. Cognitive networks, in contrast to cognitive radios, are characterized by their \emph{network-level} intelligent and self-aware behavior. In this paper, we refer to such cognitive networks as `cognitive wireless networks' (CWNs)\footnote{We use CWNs to refer to `cognitive networks' to ensure consistent naming of the three wireless networking trends (SWNs, CWNs, and VWNs) proposed in section \ref{sec:threetrends}.}. CWNs employ a cognition loop (as can be seen in figure \ref{fig:ooda}) to observe the environment, orient itself and thereafter decide/ plan to arrive at the best decision according to the networking/ user/ and application context.

While DSA is the most popularly cited application of CRNs, developing network-level intelligence in CRNs will enable numerous other applications---including the ability to reprogram itself optimally according to the network conditions.

In previous CRN research, it has been observed that PHY and MAC layers offer many ``\emph{knobs}'' that can be tweaked to optimize performance which may be measured through some ``\emph{meters}''. In \cite{fette2009cognitive}, many examples of knobs and meters at the PHY and MAC layers have been provided. Since CRNs operate in dynamic, often unknown, conditions, configuring the knobs optimally is not a trivial problem. Various AI based techniques have been proposed in literature to assist CRNs in their quest of performing autonomous optimal adaptations in such settings \cite{he2010survey} \cite{qadir2013artificial}. Apart from artificial intelligence, CRN also borrows techniques and tools from various other fields such as game theory, control theory, optimization theory, metaheuristics, etc. \cite{haykin2005cognitive}.

While device-level reconfiguration capabilities (e.g., SDRs and CRs) and network-level reconfiguration capabilities (e.g., SDN) will undoubtedly be a big part of future programmable wireless networking, the resulting programmable wireless architecture will still not be fully automated unless AI techniques are incorporated into the core of the framework. In addition to programmable data plane and programmable control plane, both offering various useful abstractions, future wireless programmable networking also requires a ``\emph{knowledge plane}'' \cite{clark2003knowledge}. Since CRNs inherently embody AI techniques with wireless communications, it seems natural to explore using CRs, along with the capabilities of SDN and SDRs, to provide mechanisms for implementing the knowledge plane of future programmable wireless devices.

In future work, the hybrid use of SDN and CRN technology could plausibly lead to a more powerful programmable wireless networking paradigm. While the CogNet project \cite{raychaudhuri2006cognet} did propose an architectural model of separated control and data plane with an extensible global control plane controlling the separated data planes through an API---which is similar in spirit to the SDN architecture---no concrete proposal has followed this initial conception. This area is ripe for further exploration to exploit the best of CRN and SDN worlds.

We note here that CWNs are \emph{autonomously self-programmed networks}, i.e., CWNs incorporate the ability of autonomously adapting, or \emph{programming}, itself so that operation parameters are optimized to fulfill the desired goals of performance. This conception of programmability in CWNs is distinct from the traditional view of programmability (which also applied to SWNs) which implicitly assumes non-autonomous programming. Future programmable wireless networking will arguably employ both autonomous and non-autonomous programming to reap the benefits of both approaches.

In the following subsections, we will introduce some autonomous programming applications, or adaptive features, of CWNs. We will focus on the features of DSA, co-existence facilitation, and integration with cloud technology.

\vspace{2mm}
\subsubsection{Dynamic spectrum access (DSA)-based CWNs}

An important adaptive feature of CWNs is in the reconfiguration of operating frequency of a secondary user (SU) in a DSA networks. This depends critically on spectrum sensing (which is performed to detect the presence of primary users, or PUs) which is used to ensure that incumbent licensed users, represented by the PUs, are not interfered with. In certain cases, spectrum sensing can be avoided and a database lookup specifying the activity pattern of the PU suffices \cite{zhao2007survey}.

\vspace{2mm}
\subsubsection{Coexistence-facilitating CWNs}

Many IEEE standards (such as IEEE 802.11, 802.15, and 802.16) incorporate some basic cognitive radio functionality such as dynamic frequency selection (DFS) and power control (PC) which facilitate coexisting networks sharing the same frequency \cite{sherman2008ieee}.

\vspace{2mm}
\subsubsection{Cloud-based CWNs}

There have been a few proposals for utilizing cloud technology to improve the performance in CWNs. In particular, cloud technology has been proposed for use with CWNs for cooperative spectrum sensing in the TV white spaces \cite{ko2011cooperative} and for a fast processing of vast volumes of data \cite{ge2010cognitive}. In the future, CWNs will perceivably leverage cloud technology increasingly to exploit its scalable computation capability along with its inherent programmability.

\subsection{Trends 3: Virtualizable Wireless Networks (VWNs)}
\label{sec:VWN}

Virtualization has transformed both the operational efficiency and the economics of the compute industry, and more recently, the datacenter environment. With the growing role of virtualization in networking, it is highly likely that future programmable wireless networks will be virtualization based. An important application ``pull'', or use-case, of VWNs in general is the convenience of supporting multiple tenants on shared infrastructure. An analogue of this use case from the service provider's perspective is the need to support multiple virtual network operators (VNOs) or mobile VNOs (MVNOs) when talking in the perspective of mobile carrier networks \cite{sachs2008virtual}. In the following, we shall discuss the application of virtualization in four environments: \emph{i)} WLANs, \emph{ii)} mobile carrier networks, \emph{iii)} clouds, and \emph{iv)} CRNs. For a more comprehensive discussion of wireless virtualization, the interested reader is referred to a specialized book on this topic \cite{wen2013wirelessbook}.

\vspace{2mm}
\subsubsection{WLAN-based VWNs}

With the widespread use of IEEE 802.11 WLANs (Wi-Fi)---and the pervasive commissioning of Wi-Fi hotspots in campuses, offices, business centers, airports, shopping centers, etc.----the wireless signal is almost ubiquitously available. There has been a lot of interest in exploiting this common infrastructure to support multi-tenant and multi-provider environments. The concept of `slices' proposes to provide virtualized environment that runs on top of common shared infrastructure. The general area of network virtualization (NV) is explored in depth in \cite{chowdhury2010survey}, and the interested reader is referred to this paper, and the references therein, for more details. Some prominent contributions that have proposed virtualization for WLANs include:  wireless virtualization on commodity 802.11 hardware \cite{smith2007wireless}, the use of virtual access points (VAPs) \cite{hamaguchi2010framework}, virtual Wi-Fi \cite{xia2011virtual}, and building multi-purpose access point (MPAP) virtualization architecture \cite{he2010mpap}.

\vspace{2mm}
\subsubsection{Cellular Mobile VWNs}

With the mobile traffic increasing exponentially \cite{raychaudhuri2012frontiers}, mobile carrier wireless networks is an attractive setting for wireless virtualization. Various works, focusing on mobile carrier VWNs, have exploited virtualization technology, with some sample works being Costa et al.'s work on RAN Virtualization \cite{costa2013radio}, Bhanage's work on WiMAX base station virtualization \cite{bhanage2010virtual}, and Zaki et al.'s work on LTE virtualization \cite{zaki2010lte}.

\vspace{2mm}
\subsubsection{Cloud-based VWNs}

Cloud technology has recently been used with VWNs to provide its scalability and service benefits in such environments. The CloudMAC project has proposed virtualizing access points (APs) in a datacenter \cite{vestin2013cloudmac}. Vassilaras et al. present an approach in \cite{vassilaras2011wireless} to provide wireless networking as a service (in utility computing style associated with cloud computing). Other VWN projects that have utilized cloud computing include the wireless network cloud (WNC) project \cite{lin2010wireless}. Due to their popular usage, and associated benefits, cloud technology has also been proposed for use with SDRs \cite{gomez2012resource} and in CRs \cite{ge2010cognitive}. A recent trend in cloud computing is mobile cloud computing in which clients use mobile wireless devices to perform computations in the cloud. A detailed survey of mobile cloud computing is provided in \cite{dinh2011survey}.

\vspace{2mm}
\subsubsection{CRN-based VWNs}

Relatively less work has been done on CRN-based VWNs. In \cite{tan2012enable}, Tan et al. have presented a novel spectrum virtualization layer,  that runs directly below the wireless PHY layer, that presents a seamless interface to the upper layers while allowing dynamic spectrum access (DSA).

\section{Open Research Issues and Challenges}
\label{sec:openquestions}

Research in programmable wireless networks has indeed gained momentum as highlighted in this paper. However, many issues still remain to be resolved in order to fully realise the potential benefits associated with this paradigm. We highlight a few important research issues in this domain.

\subsection{Building Software Defined Cognitive Wireless Networks}

An initial promise of software defined radio (SDR) was seamless interworking with a plethora of technologies through software defined adaptations. The vision of CRNs, on the other hand, has evolved from the foundations of SDRs and aims to provide users with seamless holistic experience that integrates potentially heterogeneous technologies. The interplay of cognitive radios with the SDN architecture appears to be a viable and promising hybrid technology that can be used to create programmable wireless networks. We refer to this hybrid technology as software defined cognitive wireless networking (SCWN). Numerous interesting use-cases can plausibly emerge if we synergize the mainly centralized operational paradigm of SDNs with the mainly distributed operational paradigm of CRs. While the emphasis of SDN architecture has been on the separation of control and data planes, it is worth exploring if a combined SDN and CR architecture can help realize the vision of programmable wireless networks having a `knowledge plane' as envisioned by Clark et al. \cite{clark2003knowledge}.

\subsection{Development of Wireless Specific Network APIs}

The utility of any programming paradigm depends greatly on the abstractions available and the standardization of APIs. For the vision of programmable wireless networking to become established, it is important that there is progress in developing useful network APIs offering sophisticated high-level abstractions for wireless networking. In the SDN community, while numerous southbound APIs have been proposed, there is a lack of clarity and consensus about what abstractions and interface a northbound API should expose. While OpenFlow, an example southbound API, has become wildly popular, it is quite primitive in its functionality \cite{foster2013languages}---the metaphor of Assembly language programming is often used to describe direct OpenFlow programming. To increase the rate of innovation, it is important that higher level languages are developed that can be used by network programmers through a high-level standardized interface. In this regard, more work is required for both southbound and northbound API. Since a network application programmer will interface with the controller through a northbound API, quick consensus on the development of an effective northbound network API is of paramount importance.

\subsection{Integrating Wireless and Cloud Technologies}

The paradigm of cloud computing---which is itself based on web technology, programmability through APIs, and virtualization---is likely to play a big role in future wireless programmable networks. There is already a lot of work on integration of SDN and cloud technology \cite{Neutron}. Future designers of programmable wireless devices will be well-served by exploiting the performance and scalability advantages offered by cloud computing in their designs. Already, there has been significant work in incorporating cloud technology into existing frameworks, and this trend looks set to continue well into the future.

\subsection{Wireless Internet of Things}

While traditionally the Internet communication paradigm has revolved around human consumption of Internet services, it is envisioned that in the future, networking will create many novel services through machine-to-machine communication by creating an Internet of things \cite{atzori2010internet}. In particular, the convenience of untethered mobile communication facilitated by wireless communication can create a future wireless Internet of things. This is a potential future research area envisioned to have a significant impact on the community and the way stack holders interact with the services provided by Internet.

\subsection{Balancing Centralized and Distributed Paradigms}

SDNs have proposed a separation of the control plane and data plane with the control logic placed on a separated controller. Pragmatic concerns about performance and security have dictated that this controller be implemented as a distributed system. Hence, although the controller is `logically centralized', it is implemented as a distributed system---this has led to coining of the awkward term ``logically centralized control''. This draws our attention to the perennial tension between distributed and centralized control. The Internet's community has traditionally favored the distributed control paradigm due to its scalability. However, architectural ossification and inflexible network control has led through a rethink to the centralized SDN paradigm. Like the pre-SDN era, not all tasks can be, or should be, exclusively centralized or distributed. The modern shift to a centralized paradigm is sometimes codified in the mantra, ``centralize what you can, distribute what you must''. Finding the right balance between centralized and distributed control is an important fundamental design choice which needs careful evaluation. Also, it is important to address the scalability and performance concerns associated with the centralized control to make it a viable practical architecture. Future wireless networks will have to seamlessly manage the delicate balance between the centralized and distributed control paradigms of current technologies---such as WiMAX and Wi-Fi---and the centralized aspects and distributed aspects of future network architecture such as SWNs and CWNs, respectively.

\section{Conclusions}
\label{sec:conclusion}

In this paper, we have provided a general overview of architectural techniques useful for building next-generation programmable wireless networks. We have seen that the seemingly disparate schemes of software defined radio, cognitive radio networking, software defined networking and programmable wireless processors are in fact themed on a common goal of creating flexible ``programmable wireless networks''. A self-contained tutorial is provided for these architectures followed by a detailed survey of their applications. We also proposed synergizing these technologies into newer hybrid technologies. In particular, we proposed a new framework of \emph{software defined cognitive wireless networking} which will employ both SDN and CRN principles to potentially open up new use cases. We have also highlighted important research issues in this field and identified future research work.

\bibliographystyle{ieeetr}
\bibliography{t1}

\begin{IEEEbiography}[{\includegraphics[width=1in,height=1.25in,clip]{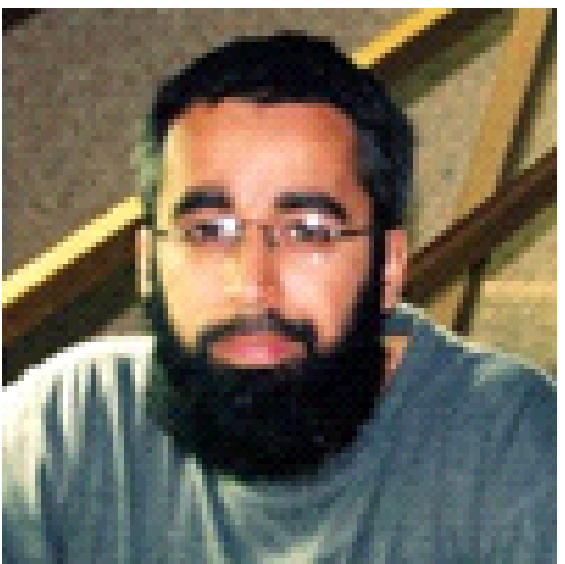}}]{Junaid Qadir} is an Assistant Professor at the School of Electrical Engineering and Computer Sciences (SEECS), National University of Sciences and Technology (NUST), Pakistan. He is also the Director of the Cognet Lab at SEECS. He completed his BS in Electrical Engineering from UET, Lahore, Pakistan and his PhD from University of New South Wales, Australia in 2008. His research interests include cognitive radio networks, wireless networks, and software defined networks.
\end{IEEEbiography}

\begin{IEEEbiography}[{\includegraphics[width=1in,height=1.25in,clip,keepaspectratio]{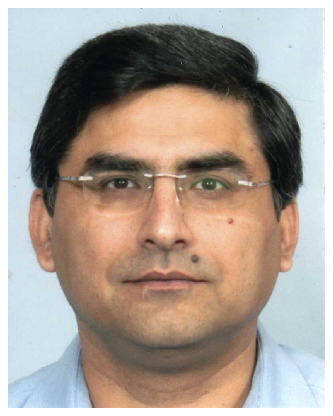}}]{Nadeem Ahmed} received the BE degree from the University of
Engineering and Technology, Lahore, Pakistan and the MS and PhD degrees
in computer sciences from University of New South Wales (UNSW), Sydney,
Australia in 2000 and 2007, respectively. He is currently an Assistant Professor at the School of Electrical Engineering and Computer Sciences (SEECS), National University of Sciences and Technology (NUST), Pakistan. His research interests include wireless sensor networks, software defined networks and mobile ad-hoc networks.
\end{IEEEbiography}

\begin{IEEEbiography}[{\includegraphics[width=1in,height=1.25in,clip,keepaspectratio]{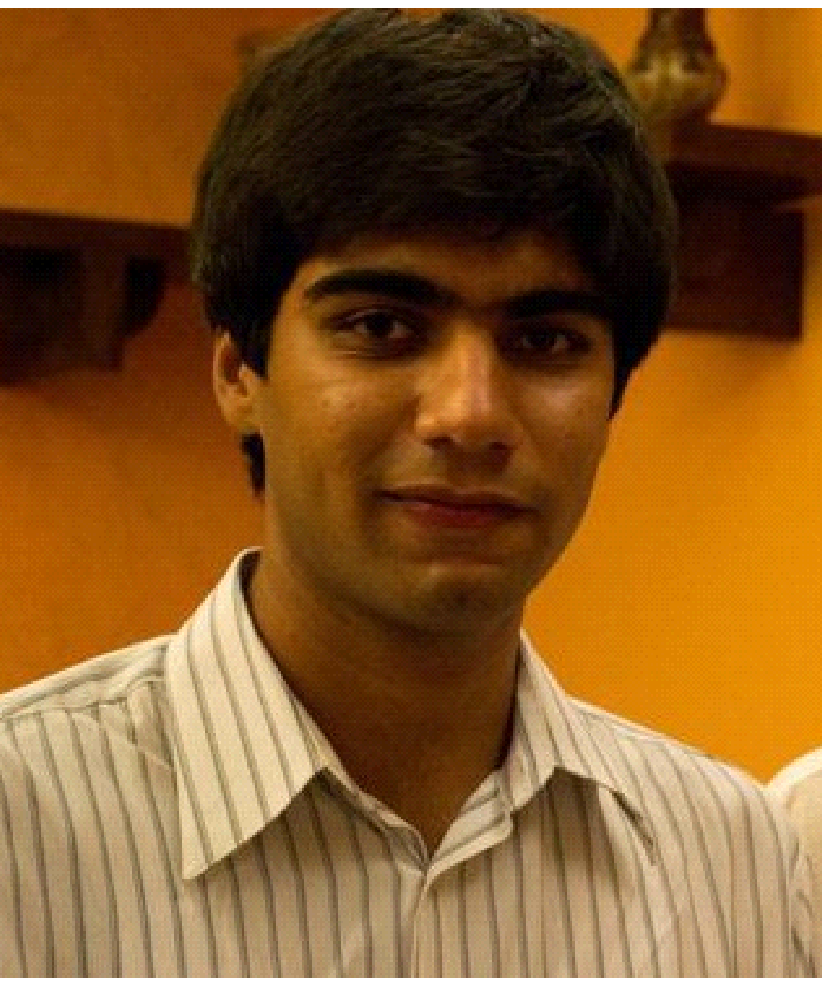}}]{Nauman Ahad} is an undergraduate Electrical Engineering student at the School of Electrical Engineering and Computer Sciences (SEECS), National University of Sciences and Technology (NUST), Pakistan. He was an internee in the Cognet research lab at SEECS, NUST, in the summer of 2013.
\end{IEEEbiography}

\end{document}